\documentclass[aip, jap, amsmath,amssymb, preprint]{revtex4-1}
\usepackage{dsfont}
\usepackage{graphicx}
\usepackage{dcolumn}
\usepackage{bm}
\usepackage[mathlines]{lineno}
\usepackage[utf8]{inputenc}
\usepackage[T1]{fontenc}
\usepackage{mathptmx}
\usepackage{xcolor}
\usepackage{hyperref}
\usepackage{booktabs}
\usepackage[version=4]{mhchem}
\usepackage{subfigure}

\usepackage{todonotes,tcolorbox}
\definecolor{yellow2}{HTML}{FADF63}

\newcommand{\mycomment}[1]{}

\begin{document}


\title{Elastic moduli and thermal conductivity of quantum materials at finite temperature}
\author{Dylan A. Folkner}
\affiliation{Department of Chemistry, University of California, Davis, Davis, CA 95616, USA}%
\author{Zekun Chen}
\affiliation{Department of Chemistry, University of California, Davis, Davis, CA 95616, USA}%
\author{Giuseppe Barbalinardo}
\affiliation{Department of Chemistry, University of California, Davis, Davis, CA 95616, USA}%
\author{Florian Knoop}
\affiliation{Department of Physics, Chemistry and Biology, Link{\"o}ping University, 581 83 Link{\"o}ping, Sweden}
\author{Davide Donadio}
\email{ddonadio@ucdavis.edu}
\affiliation{Department of Chemistry, University of California, Davis, Davis, CA 95616, USA}%

\date{\today}
\newcommand{\angstrom}{\text{\normalfont\AA}}
\begin{abstract}
We describe a theoretical and computational approach to calculate the vibrational, elastic, and thermal properties of materials from the low-temperature quantum regime to the high-temperature anharmonic regime. 
This approach is based on anharmonic lattice dynamics and the Boltzmann transport equation. It relies on second and third-order force constant tensors estimated by fitting temperature-dependent empirical potentials (TDEP) from path-integral quantum simulations with a first-principles machine learning Hamiltonian. 
The temperature-renormalized harmonic force constants are used to calculate the elastic moduli and the phonon modes of materials. Harmonic and anharmonic force constants are combined to solve the phonon Boltzmann transport equation to compute the lattice thermal conductivity. 
We demonstrate the effectiveness of this approach on bulk crystalline silicon in the temperature range from 50 to 1200~K, showing substantial improvement in the prediction of the temperature dependence of the target properties compared to experiments. 

\end{abstract}

\maketitle

\section{Introduction}

The viability of materials for various technologies, including electronics, and energy storage, harvesting, and conversion, depends on their mechanical and thermal properties. At the microscopic level, these properties are dictated by lattice vibrations, also called phonons. 
Specifically, phonons are the main thermal energy carriers in insulators and semiconductors, whereas their interaction with electrons is the primary limiting factor of thermal transport in metals. 

Lattice dynamics (LD) is the standard numerical approach to compute the phonon properties both in crystalline materials and in amorphous solids  because it naturally includes statistical quantum effects.\cite{phyphononsrivastava,broido_lattice_2005,broido_intrinsic_2007,ravichandran_unified_2018, mcgaughey2019phonon,simoncelli_unified_2019,isaeva2019modeling,barbalinardo2020efficient}  
LD relies on the Taylor expansion of the interatomic potential, which can be either semi-empirical or computed by first principles, usually by density functional theory (DFT). 
The second-order term in the expansion gives the normal modes and their frequencies, thus giving access to the harmonic properties of the system, such as phonon dispersion relations, group velocities, heat capacity, and elastic constants. 
Higher-order terms account for anharmonic phonon-phonon scattering events, and they are necessary to calculate the spectral linewidths and the thermal conductivity ($\kappa$), which is obtained by solving the Boltzmann transport equation (BTE).\cite{omini_beyond_1996} 
These terms are normally truncated at the third order, but four-phonon processes have been recently found essential to accurately estimate the thermal conductivity of certain materials, such as \ce{BAs}.\cite{feng_four-phonon_2017, broido_anharm, anharm_zn_blende_2020, hellman_perovskites_2020, han_fourphonon_2022}

The second and higher-order terms in the expansion of the potential energy, also identified as interatomic force constants (IFC), are computed as zero-temperature derivatives of the potential either by finite differences or by perturbation theory.\cite{baroni_phonons_2001, paulatto_anharmonic_2013} 
The harmonic IFC can be exploited in the standard quasiharmonic approximation to compute thermodynamic properties.\cite{baroni_thermal_2010, fultz_vibrational_2010}

However, this implementation of LD is viable only for systems that are stable or meta-stable at zero temperature, i.e. their harmonic IFC matrix does not exhibit negative eigenvalues which would correspond to imaginary frequencies. Additionally, even for stable systems temperature effects are only accounted for in the solution of the phonon population terms appearing in the solution of the BTE. 

The need to overcome these fundamental issues has led to the development of various methods in which IFCs at finite temperatures are inferred from molecular dynamics simulations\cite{campana_practical_2006, kong_implementation_2009, hellman2013temperature} or self-consistent stochastic sampling of atomic displacements.\cite{souvatzis_entropy_2008, errea_anharmonic_2014, monacelli_pressure_2018, van_roekeghem_anomalous_2016, fransson_efficient_2020}

These methods have been implemented in several open-source codes, including SSCHA,\cite{monacelli_stochastic_2021} QSCAILD,\cite{van_roekeghem_quantum_2021} HiPHive,\cite{brorsson_efficient_2022} ALAMODE,\cite{tadano_anharmonic_2014} TDEP\cite{knoop2024tdep} and the ``user-phonon" package in LAMMPS.\cite{kong_implementation_2009}



In this Tutorial Article, we illustrate a state-of-the-art workflow to compute the temperature-renormalized vibrational, elastic, and thermal transport properties of crystals by anharmonic LD at finite temperatures integrating nuclear quantum effects through Feynman's path integral. 
The application of this workflow to crystalline silicon shows that even for such a simple system, nuclear quantum effects and finite-temperature displacements must be considered to obtain accurate estimates of the thermal conductivity, especially at low temperatures. 
To achieve convergence in the sampling of the forces used to compute the IFCs, we exploit a machine-learning neuroevolution potential (NEP), consisting of a single-layer neural feed-forward network model fitted on a DFT database with a natural evolutionary algorithm.\cite{fan2021neuroevolution} The forces extracted from path-integral molecular dynamics (PIMD) simulations implemented in the GPUMD code,\cite{fan2022gpumd} are used to compute IFCs by a multi-order force constant fitting through the temperature-dependent effective potential (TDEP) method.\cite{knoop2024tdep}
Finally, the IFCs are used to compute the phonon properties, elastic constants, and thermal conductivity at each temperature as implemented in  $\kappa$ALDo.\cite{barbalinardo2020efficient} 
Each step of this workflow takes full advantage of GPU acceleration for enhanced efficiency.

In the next Section, we illustrate the theory of anharmonic lattice dynamics, the background of the methods used to compute IFCs at finite temperatures, and the calculation of mechanical and thermal properties. 
In Section III, we describe the computational workflow, and in Section IV we report the results for bulk crystalline silicon, with a critical discussion of how they differ from those obtained by standard approaches. 


\section{Theoretical background}

\subsection{Anharmonic Lattice Dynamics}
\label{sec:ALD}

For a crystal with periodic symmetry, we define atomic coordinate $\alpha$ of atom $i$ in cell $n$ at time $t$ as:
\begin{equation}\label{eqn:r_of_t}
    r^{(n)}_{i\alpha}(t) = r_{i\alpha}^{(n)}(0) + u^{(n)}_{i\alpha}(t), 
\end{equation}
where \(r_{i\alpha}^{n}(0)\) are the equilibrium positions, and \(u^{n}_{i\alpha}(t)\) is the time-dependent displacement. Expanding the potential energy, \(U\), at equilibrium and truncating up to 3rd order:
\begin{equation}\label{U_expans}
    \begin{split}
        U &\approx U_{0} + \frac{1}{2}\sum_{ni\alpha mj\beta}\left.\frac{\partial^{2}U}{\partial u^{n}_{i\alpha}\partial u^{m}_{j\beta}}\right|_{0}u^{n}_{i\alpha}u^{m}_{j\beta} \\
        &+ \frac{1}{6}\sum_{ni\alpha mj\beta hk\gamma}\left.\frac{\partial^{3}U}{\partial u^{n}_{i\alpha}\partial u^{m}_{j\beta}\partial u^{h}_{k\gamma}}\right|_{0}u^{n}_{i\alpha}u^{m}_{j\beta}u^{h}_{k\gamma} \\ 
        &\equiv U_{0} + \frac{1}{2}\sum_{ni\alpha mj\beta}C_{i\alpha j\beta}^{nm}u^{n}_{i\alpha}u^{m}_{j\beta} \\ 
        &+ \frac{1}{6}\sum_{ni\alpha mj\beta hk\gamma}C_{i\alpha j\beta k\gamma}^{nmh}u^{n}_{i\alpha}u^{m}_{j\beta}u^{h}_{k\gamma},
    \end{split}
\end{equation}
where the first derivative term is null for systems at local minimum, and \(C_{i\alpha j\beta}^{nm}\) and \(C_{i\alpha j\beta k\gamma}^{nmh}\) are the second and third-order interatomic force constants: IFC2 and IFC3, respectively. 

By further assuming the form of the displacement functions are plane waves dependent on wavevector $\pmb{q}$ and frequency $\omega_\mu$, it can be proven that
\begin{equation}\label{dyn_eq}
    \omega_{\mu}^{2}(\pmb{q})\eta_{i\alpha}^{\mu}(\pmb{q}) = \sum_{j\beta}D_{i\alpha j\beta}(\pmb{q})\eta_{j\beta}^{\mu}(\pmb{q}),
\end{equation}
where \(D_{i\alpha j\beta}(\pmb{q})=\sum_{m}\sqrt{M_i^{-1}M_j^{-1}} C_{i\alpha j\beta}^{0m}e^{-i\pmb{q}\cdot\pmb{R}_{m}}\) is the mass reduced Fourier transform of the interatomic force constant matrix \(C_{i\alpha j\beta}^{nm}\), referred to as the dynamical matrix, \(\omega_{\mu}^{2}(\pmb{q})\) is the square frequency of vibrational mode \(\mu\) at wavevector \(\pmb{q}\), and \(\eta_{j\beta}^{\mu}(\pmb{q})\) are the phonon eigenvectors at the \(j^{th}\) atom in the direction of coordinate \(\beta\) at the wavevector \(\pmb{q}\). Thus, our system dynamics can be described with the dynamics of the quantum harmonic oscillators of frequencies \(\omega_{\mu}\), and our system is entirely determined by these force constants, \(C_{i\alpha j\beta}^{nm}\)\cite{barbalinardo2020efficient, doveintrolattdyn, phyphononsrivastava}. Vibrations are quantized and are referred to as phonons, which do not possess an intrinsic spin and whose statistics can be modeled by the Bose-Einstein statistics. Applying the Bose-Einstein equation for the average occupancy of an oscillator, the average occupation for mode \(\mu\) at a given temperature, \(T\), is represented by,
\begin{equation}\label{bose_ein}
    n_{\mu} \equiv \bar{n}(T; \omega_{\mu}) = \frac{1}{e^{\hbar\omega_{\mu}/k_{B}T} - 1}.
\end{equation}

The harmonic approximation, i.e. the truncation of Eq.~\ref{U_expans} at the second order, not only provides the phonon dispersion relations $\omega(\pmb{q})$ but also macroscopic and thermodynamic properties of materials, such as the elastic moduli, thermal expansion, and lattice heat capacity.\cite{Ashcroft76}
Conversely, to compute the thermal conductivity, it is necessary to consider phonon-phonon scattering mechanisms that are described as a first approximation by the cubic anharmonic term in Eq.~\ref{U_expans}.  

\subsection{Interatomic Force Constants at Finite Temperature}

An accurate estimate of IFC2 and IFC3 tensors is essential to characterize the elastic and thermal properties of materials. 
In Eq.~\ref{U_expans} these tensors are defined as the second and third derivatives in the Taylor expansion of the potential energy.

This perturbative approach implies that the system is considered in its potential energy minimum at zero temperature. 
The temperature dependence enters only in the statistical distribution of phonons occupations in Eq.~\ref{bose_ein} when one computes thermodynamic or transport properties. 

This approach limits the application of LD to systems that are stable or meta-stable at zero temperature, thus excluding temperature-stabilized polymorphs, and cannot account for the temperature-induced renormalization of the vibrational frequencies, which can be substantial in strongly anharmonic materials.\cite{zheng_anharmonicity_2022,SnSe_anharmonicity} 

To overcome these limitations, approaches have been developed stemming from the self-consistent harmonic approximation (SCHA).\cite{souvatzis_entropy_2008, errea_anharmonic_2014, tadano_self-consistent_2015, hellman_scha_2017, ravichandran_unified_2018, monacelli_stochastic_2021}

In this class of methods, the normalized IFC2 tensor is represented as the curvature of the variational Gibbs free energy mapped on an effective quadratic Hamiltonian, sampled on a multivariate Gaussian distribution. When the variational minimum is obtained, the same sampling provides higher-order renormalized force constants.

An alternative class of methods relies on sampling the canonical distribution through classical molecular dynamics (MD) or Monte Carlo.\cite{campana_practical_2006, hellman2013temperature} 
In the temperature-dependent effective potential (TDEP) approach,\cite{hellman2013temperature} the harmonic and higher-order IFCs are fitted over a set of frames from MD trajectories at the desired temperature. Higher-order IFCs are fitted on the residual forces from the previous order. 
While TDEP was originally developed as an empirical method, it recently found a solid foundation in the Mori-Zwanzig projector formalism of linear response and mode-coupling theory.\cite{castellano2023mode} 
In the same work, Castellano {\it et al.} generalize the use of TDEP to path-integral molecular dynamics (PIMD) to obtain a quantum mechanical sampling of the system without assuming multivariate Gaussian distributions. 
As the theory is derived in Ref.~\onlinecite{castellano2023mode}, here we outline the practical aspects of the calculation of the IFCs. 

This consists of fitting the interatomic force constants through a linear regression on the forces $\left(f_{i}\right)$ and the residuals of the forces $\left(\delta f_{i}\right)$\cite{hellman2011lattice, knoop2024tdep, castellano2023mode}. For simplicity of notation, we choose to reindex the displacements and IFCs such that $C^{nm}_{i\alpha j\beta}\rightarrow C_{ij}$. Then, by defining the IFCs as sums of Kubo correlations:
\begin{equation}\label{kubo_ifcs_2}
    C_{ij} = -\sum_{k}\frac{\left(u_{k},f_{i}\right)}{\left(u_{j},u_{k}\right)},
\end{equation}
\begin{equation}\label{kubo_ifcs_3}
    C_{ijk} = -\frac{1}{2}\frac{\sum_{lm}\left(u_{l}u_{m},\delta f_{i}\right)}{\sum_{lm}\left(u_{j}u_{k},u_{l}u_{m}\right)},
\end{equation}
they can be calculated over a PIMD run by averaging the displacements $\left(\bar{u}_{i}(t_{n})\right)$ and forces $\left(\bar{f}_{i}(t_{n})\right)$ over the beads, which is an independent classical trajectory, and numerically integrating over time such that Eq. ~\ref{kubo_ifcs_2} and ~\ref{kubo_ifcs_3} become:
\begin{equation}\label{ave_ifc_2}
    C_{ij} = -\frac{\sum_{n}\sum_{k}\bar{u}_{k}(t_{n})\bar{f}_{i}(t_{n})}{\sum_{n}\sum_{k}\bar{u}_{j}(t_{n})\bar{u}_{k}(t_{n})},
\end{equation}
\begin{equation}\label{ave_ifc_3}
    C_{ijk} = -\frac{1}{2}\sum_{n}\frac{\sum_{lm}\bar{u}_{l}(t_{n})\bar{u}_{m}(t_{n})\delta\bar{f}_{i}(t_{n})}{\sum_{lm}\bar{u}_{j}(t_{n})\bar{u}_{k}(t_{n})\bar{u}_{l}(t_{n})\bar{u}_{m}(t_{n})}.
\end{equation}
The IFCs can be found through a linear regression:
\begin{equation}\label{reg_ifc_2}
    C_{ij}u_{j} = f_{i},
\end{equation}
\begin{equation}\label{reg_ifc_3}
    C_{ijk}u_{j}u_{k} = \delta f_{i},
\end{equation}
\begin{equation}\label{reg_ifc_4}
    C_{ijkl}u_{j}u_{k}u_{l} = \delta^{2} f_{i},
\end{equation}
where we define higher order residuals $\left(\delta^{n}f_{i}\right)$ as:
\begin{equation}\label{n_residual}
    \delta^{n}f_{i} = f_{i} - \sum_{j=2}^{n+1}f_{i}^{(j)},
\end{equation}
\begin{equation}\label{poly_force}
    f_{i_{1}}^{(j)} = C_{i_{1}i_{2}...i_{j}}u_{i_{2}}u_{i_{3}}...u_{i_{j}}.
\end{equation}
and the $j$ order IFCs can be calculated by fitting the average displacements over the PIMD beads to the difference of the forces with the $(j-1)$ order polynomial force.

It is important to observe that this framework does not require PIMD when nuclear quantum effects are negligible. At a high enough temperature, the sample can be taken from a classical MD calculation, equivalent to a 1-bead PIMD simulation. 
Alternatively, sampling can be achieved through iterative multivariate Gaussian sampling and fitting in a SCHA framework, which is mathematically equivalent to the SSCHA method.\cite{errea_first-principles_2013, errea_anharmonic_2014, monacelli_stochastic_2021}
However, Gaussian sampling can cause artificial biasing toward a more harmonic distribution. In general, while computationally expensive, PIMD sampling offers the most general way to obtain the correct quantum distributions and is the best method to fit the IFCs gauging nuclear quantum effects\cite{knoop2024tdep, castellano2023mode}.
These effects often have an important role in determining the vibrational properties of materials at low temperatures, with an impact not only on thermal transport but also on superconductivity and, in general, the phase diagram of quantum materials.\cite{errea_first-principles_2013, errea_high-pressure_2015, pickard_superconducting_2020}



\subsection{Path Integral Molecular Dynamics}

Path-integral molecular dynamics (PIMD) is a computational method to sample quantum mechanical distributions exploiting the isomorphism between quantum theory and classical statistical mechanics of polymer rings in imaginary time.\cite{FeynmanPI}
The derivation of Feynman path integral formalism and its implementation in several different forms is provided in textbooks and research articles,\cite{TuckermanText, tuckerman1996efficient, ceriotti2010efficient, rpmdcraig} and it is not a topic for this tutorial article. 
The essential knowledge is that through PIMD the canonical partition function of a system of $N$ quantum particles can be sampled approximately by representing each particle as a ring polymer of $n$ beads connected by harmonic springs: 
\begin{equation}
    \begin{split}
        Z &= Tr\left[e^{-\beta\hat{H}}\right]\\ &\approx \frac{1}{(2\pi\hbar)^{nN}}\int\int d^{nN}\pmb{q}d^{nN}\pmb{p}e^{-\frac{\beta}{n}H_{n}},
    \end{split}
    \label{eq:ZPIMD}
\end{equation}
where \(H_{n}\) is path integral Hamiltonian defined as:
\begin{equation}\label{bead_Hn}
    H_{n}\equiv H_{n}^{0} + V_{n},
\end{equation}
\begin{equation}\label{bead_HO}
    H_{n}^{0} = \sum_{i=1}^{N}\sum_{j=1}^{n}\left(\frac{(p_{i}^{j})^{2}}{2m_{i}} + \frac{1}{2}\omega_{n}^{2}\left(q_{i}^{j}-q_{i}^{j-1}\right)^{2}\right)
\end{equation}
\begin{equation}\label{bead_V}
    V_{n} = \sum_{j=1}^{n}V(\pmb{q}^{j}),
\end{equation}
where \(q_{i}^{j}\) and \(p_{i}^{j}\) are \(i^{th}\) canonical position and momentum coordinates of the \(j^{th}\) bead\cite{ceriotti2010efficient, craig2004quantum, pimdsupp}.

For fitting IFCs at the quantum level, the advantage of using PIMD over other sampling techniques that assume a Gaussian probability distribution of displacements is that it can naturally account for large deviations from harmonic potentials, as in the case of solid helium.\cite{castellano2023mode}
Additionally, PIMD can be performed at constant pressure by coupling the dynamics with a barostat.\cite{martyna_molecular_1999} 
This allows us to obtain the equilibrium density at any temperature and compute the thermal expansion without any further approximation. 

The isomorphism represented in Eq.~\ref{eq:ZPIMD} is exact in the limit of $n\rightarrow\infty$. 
In practical applications, the number of beads needed to achieve convergence depends on the material and the temperature. Materials with light atoms and low temperatures require larger $n$. Convergence in the number of beads must be tested carefully. All PIMD simulations were conducted using GPUMD-3.8\cite{fan2022gpumd}.



\subsection{Machine Learning Potentials}

The high costs of PIMD sampling and the need to simulate sufficiently large supercells make the direct use of first-principles DFT impractical.  
To efficiently sample the free energy surface at different temperatures, we employ a machine-learning potential (MLP) trained on a suitable dataset of energies and forces computed by DFT. 
MLPs have been used extensively to predict vibrational and thermal properties of both crystalline and disordered materials with very promising results since the early applications on phase change and thermoelectric compounds.\cite{sosso_thermal_2012,campi_electron-phonon_2015, bosoni_atomistic_2020,mangold_transferability_2020} 
An exhaustive list of thermal transport studies with MLPs is provided in Ref.~\cite{dong_molecular_2024}. 

Here we chose the neuroevolution potential (NEP) approach,\cite{fan2021neuroevolution} which consists of a feed-forward neural network with 2-body \((d_{n}^{i})\), 3-body \((d_{nl}^{i})\), 4-body \((d^{i}_{nl_{1}l_{2}l_{3}})\), and 5-body descriptors \((d^{i}_{nl_{1}l_{2}l_{3}l_{4}})\). 
The native implementation of NEP on graphical processing units (GPUs) in the GPUMD code\cite{fan2022gpumd} provides a substantial computational advantage, making this approach very efficient for running PIMD with large $n$ \cite{ying_highly_2024}.

The dataset used to fit the NEP model contained 2475 structures for crystalline silicon with varying deformations, vacancies, bond numbers, and various polymorphs, including amorphous silicon\cite{bartok2018machine}. DFT calculations were performed using the Perdew-Wang (PW91) generalized gradient approximation functional.\cite{perdewwang91}

\subsection{Thermal Expansion Coefficients in the Quasi-Harmonic Approximation}

The volumetric coefficient of thermal expansion, \(\alpha_{V}(T)\), is defined as the rate of change of the volume relative to the current volume, through the equation:
\begin{equation}\label{therm_exp}
    \alpha_{V} (T) = \frac{1}{V}\left(\frac{\partial V}{\partial T}\right)_{p}.
\end{equation}
In a cubic crystal, such as bulk silicon, thermal expansion is isotropic and can be equivalently characterized by the linear thermal expansion coefficient \(\alpha_{L}(T)\): 
\begin{equation}\label{lin_therm_exp}
        \alpha_{V}(T) = 3\alpha_{L}(T),
\end{equation}

Here, we recommend computing equilibrium densities at any temperature by constant-pressure PIMD. It is useful to compare this approach to a commonly used method to compute thermal expansion in crystals dubbed Quasi-Harmonic Approximation (QHA).\cite{biernacki_negative_1989, fleszar_first-principles_1990, pavone_old_2001, baroni_thermal_2010, togo2010first, ritz2019thermal} 
In QHA, the approximate free energy of a crystal is expressed in terms of the harmonic phonon energies, $E_{ph}(T)$, and entropic contributions, $S_{ph}(V,T)$:
\begin{equation}\label{helmholtz_qha}
    F(V, T) = E_{el}(V) + E_{ph}(V,T) - TS_{ph}(V,T),
\end{equation}
where $E_{el}(V)$ is the zero-temperature energy of the crystal, and
\begin{equation}\label{ph_energy}
    E_{ph}(T) = \sum_{\mu}\hbar\omega_{\mu}\left(n_{\mu} + \frac{1}{2}\right),
\end{equation}
which reduces to the zero-point energy contribution for $n_\mu=0$.
The harmonic entropic term is:
\begin{equation}
\label{eq:entropyQHA}
  S_{ph} = -k_B \sum_\mu \left( 1- e^{-\frac{\hbar\omega_\mu}{k_BT}} \right).
\end{equation}
The equilibrium volume at a given pressure and temperature is the volume that minimizes the Gibbs free energy:
\begin{equation}\label{Eq:GibbsQHA}
   G(P,T) = \min_V \left( E_{el}(V) + E_{ph}(V,T) - TS_{ph}(V,T) + PV \right).
\end{equation}
Whereas $F(V,T)$ is harmonic, in QHA the dependence of the phonon frequencies on the volume accounts for anharmonic effects and gives a non-zero thermal expansion coefficient at a zero-order approximation. Temperature effects are included only through the equilibrium phonon distribution function in the expression of the phonon entropy. 
\mycomment{
QHA incorporates temperature dependence through the Bose-Einstein distribution, Eq. ~\ref{bose_ein}, and the phonon energy,
\begin{equation}\label{ph_energy}
    E_{ph}(T) = \sum_{\mu}\hbar\omega_{\mu}\left(n_{\mu} + \frac{1}{2}\right),
\end{equation}
where the zero temperature phonon energy is also known as the zero-point energy (ZPE),
\begin{equation}\label{zpe}
    E_{ZPE} = E_{ph}(0) = \sum_{\mu}\frac{1}{2}\hbar\omega_{\mu},
\end{equation}
In other words, the quasi-harmonic approximation calculates \(V(T)\) by minimizing:
\begin{equation}\label{helmholtz_qha}
    F(V, T) = E_{el}(V) + E_{ph}(T) - TS(V,T),
\end{equation}
with respect to \(V\) at a given \(T\), where \(S(V,T)\) is the Von Neumann definition of entropy.} 

On the practical side, a QHA calculation consists of computing the zero-temperature energy and phonon frequencies for a grid of volumes around the equilibrium volume and using them to evaluate Equations~\ref{helmholtz_qha} and ~\ref{Eq:GibbsQHA}. 
Since all the calculations are performed at zero temperature, QHA cannot be applied to temperature-stabilized crystals and does not account for the temperature renormalization of phonon frequencies, thus leading to substantial errors for strongly anharmonic systems. In this paper, our QHA simulation utilizes 11 volumes (0 $\%$ and $\pm 5 \%$ with 1 \% increments.) with each QHA calculation performed on a $4\times4\times4$ supercell using the 8-atom conventional cell and q-point mesh of $48\times48\times48$. 

\subsection{Elastic Moduli}

In continuum mechanics, the strain tensor, \(\pmb{\epsilon}\), is a measure of the deformation relative to the original shape of the object, while the stress tensor, \(\pmb{\sigma}\), is a tensor which acts as a measure of force per area of a surface. Under small deformation, the system obeys Hooke's Law, where the stress is linearly proportional to the strain through the elastic constant tensor, \(\pmb{c}\), such that:
\begin{equation}\label{hooke_law}
    \sigma_{ij} = c_{ijkl}\epsilon_{kl},
\end{equation}
where \(i,j,k,l\in\{1,2,3\}\)\cite{dftelastic3}, the Cartesian indices. Since the stress and strain are symmetric, the tensor equation can be rewritten as a matrix equation and reindexed using Voigt notation\cite{dftelastic3},
\begin{equation}\label{hooke_voigt}
    \sigma_{i} = c_{ij}\epsilon_{j},
\end{equation}
where the stress and strain have been converted into 6-vectors, and the elastic tensor converted into a 6x6 square matrix. It can be shown that the bulk modulus is related to these elastic constants. For a cubic crystal, that relationship is:
\begin{equation}\label{bulk_mod}
    B \equiv \frac{1}{K} = \frac{c_{11}+2c_{12}}{3},
\end{equation}
where \(K\) is the isothermal compressibility\cite{elasticcubes},
\begin{equation}\label{isotherm}
    K = -\frac{1}{V}\left(\frac{\partial V}{\partial P}\right)_{T}.
\end{equation}
Similarly, to calculate Young's moduli, equations using the elastic constants can also be derived for the nonequivalent facets\cite{youngmoduli}:
\begin{equation}\label{young_111}
    E_{<111>} = \frac{3}{\frac{1}{3B} + \frac{1}{c_{44}}}
\end{equation}
\begin{equation}\label{young_110}
    E_{<110>} = \frac{2}{\frac{c_{11}}{3B(c_{11}-c_{12})} + \frac{1}{2c_{44}}}
\end{equation}
\begin{equation}\label{young_100}
    E_{<100>} = \frac{3B(c_{11}-c_{12})}{c_{11}+c_{12}}
\end{equation}
The elastic shear modulus, $G$, is instead defined as:
\begin{equation}
    G=\frac{3c_{44}+c_{11}-c_{12}}{5}
\end{equation}

As such, accurate modeling of the elastic constants of a material is necessary for understanding how the material responds to strain and material strength.

Whereas it is possible to estimate the elastic constants directly from Hooke's law, by small deformations to a crystal and computing $\sigma_{ij}$, this approach involves large uncertainties at finite temperature.\cite{sprik_second-order_1984} It is convenient to exploit Born's long wave method\cite{BornHuang} and derive the elastic constants from the dynamical matrix in Eq.~\ref{dyn_eq}, as:
\begin{equation}\label{elast_const}
    c_{ijkl} = [ik,jl] + [jk,il] - [ij,kl] + (ij,kl),
\end{equation}
where we have defined
\begin{equation}\label{brakcet}
    [ij,kl] = \frac{1}{2V}\sum_{ij}\sqrt{M_{n}M_{m}}D^{nm}_{ij,kl},
\end{equation}
\begin{equation}\label{brace}
    (ij,kl) = -\frac{1}{V}\sum_{nm}\sum_{rs}\sum_{ph}\sqrt{M_{m}M_{h}}D^{nm}_{ri,j}\Gamma^{np}_{rs}D^{ph}_{sk,l},
\end{equation}
\begin{equation}\label{Gamma}
    \Gamma^{np}_{rs} = \sum_{\mu}\frac{\eta_{nr}^{\mu}\eta_{ps}^{\mu}}{\omega_{\mu}^{2}},
\end{equation}
where \(D^{nm}_{ri,j}\) and \(D^{nm}_{ij,kl}\) are the first and second derivatives of the dynamical matrix, where the \(j\) and \(kl\) indicate the directions of the derivatives\cite{elasticborn,elasticborn2}. 
This approach is implemented in the $\kappa$ALDo package to be used with the dynamical matrix obtained by fitting PIMD simulations with the TDEP method. However, the sensitivity of this method to cell size and potential cutoffs requires independent convergence of and may necessitate larger supercells for accurate elastic constants.



\subsection{Thermal Conductivity}

Anharmonic lattice dynamics (ALD) provides a natural framework to compute the thermal conductivity of crystalline solids through the linearized BTE.\cite{phyphononsrivastava, ziman2001electrons, sparaacbte, broido_lattice_2005, mcgaughey2019phonon, lindsay2020thermal}  
In the absence of external fields and at stationary conditions, the BTE for a nonequilibrium phonon distribution \({n'}(x,T)\) reads:
\begin{equation}\label{therm_bte}
    \pmb{v}_{\mu}\cdot\nabla T\frac{\partial n'_{\mu}}{\partial T} = \left(\frac{\partial n'_{\mu}}{\partial t}\right)_{coll}.
\end{equation}
The right-hand side of the equation comprises both intrinsic phonon-phonon scattering mechanisms and extrinsic scattering with defects and boundaries.
The BTE is linearized by expanding the phonon populations to the first order around their equilibrium distribution $n^0$,
\begin{equation}\label{n_taylor}
    n_{\mu\alpha}' \approx n^0_{\mu} + \lambda_{\mu\alpha}\frac{\partial T}{\partial x_{\alpha}}\frac{\partial n_{\mu}'}{\partial T},
\end{equation}
where the term \(\lambda_{\mu\alpha}\) is the mean free path of phonon \(\mu\) in the direction of coordinate \(\alpha\) and \(n'_{\mu}\) is the non-equilibrium density of mode \(\mu\). 
We can approximate the heat current density per mode in the direction of \(\alpha\), \(j_{\mu\alpha}\), as:
\begin{equation}\label{modal_flux}
    \begin{split}
        j_{\mu\alpha} &\approx -\frac{1}{N_{q}V}\sum_{\beta} \hbar\omega_{\mu}v_{\mu\alpha}(n'_{\mu\beta}-n^0_{\mu}) \\ 
        &= -\frac{1}{N_{q}V}\sum_{\beta}c_{\mu}v_{\mu\alpha}\lambda_{\mu\beta}\frac{\partial T}{\partial x_{\beta}},
    \end{split}
\end{equation}
where \(c_{\mu} \equiv \hbar\omega_{\mu}\frac{\partial n_{\mu}}{\partial T}\) is the specific heat of the mode \(\mu\), \(N_{q}\) is the number of $q$-points, and \(V\) is the unit cell lattice volume. Thus, by summing the modal contributions to the current density and equating the result to Fourier's Law, we obtain a modal expression for the lattice thermal conductivity tensor:
\begin{equation}\label{kappa}
    \kappa_{\alpha\beta} = \frac{1}{N_{q}V}\sum_{\mu}c_{\mu}v_{\mu\alpha}\lambda_{\mu\beta},
\end{equation}
which means, to find the thermal conductivity, we need only the phonon mode specific heat \(\left(c_{\mu}\right)\), mode velocities \(\left(v_{\mu\alpha}\right)\), and mean free path \(\left(\lambda_{\mu\alpha}\right)\)\cite{doveintrolattdyn, sparaacbte, barbalinardo2020efficient}. The two former quantities are calculated from the second-order IFCs, as they are properties of the harmonic normal modes. The mean free path, instead, is calculated by inverting Eq. ~\ref{therm_bte}, which can be cast as a linear problem:
\begin{equation}\label{mfp}
    \lambda_{\mu\alpha} = \sum_{\mu'}\Gamma^{-1}_{\mu\mu'}v_{\mu'\alpha}.
\end{equation}
\(\Gamma\) is the scattering tensor, which, if we consider only three-phonon scattering processes, is calculated by projecting the third-order IFC tensor on the normal modes and applying Fermi's Golden Rule.\cite{cepellotti_thermal_2016, barbalinardo2020efficient}
The matrix inversion in Eq.~\ref{mfp} can be circumvented by solving the BTE self-consistently,\cite{sparaacbte, broido_lattice_2005} or variationally.\cite{fugallo_ab_2013} 
It is also possible to get an approximate solution by applying the relaxation time approximation, which however systematically underestimates the thermal conductivity.\cite{mcgaughey2019phonon}
The linearized phonon BTE framework is general and can include higher-order phonon scattering processes\cite{feng_quantum_2016} as well as defect and boundary scattering.\cite{tamura_isotope_1983, cepellotti_boltzmann_2017, barbalinardo_ultrahigh_2021}

The standard implementation of anharmonic lattice dynamics to compute the vibrational properties and the thermal conductivity of crystals relies on the calculation of IFCs at zero temperature, either by finite differences\cite{togo2010first} or by perturbation theory.\cite{baroni_phonons_2001, paulatto_anharmonic_2013} 
The details of this workflow are given in Ref.~\cite{mcgaughey2019phonon}
Solutions of the linearized phonon BTE are implemented in several open-source codes, including Phono3py\cite{togo2023first}, ShengBTE\cite{li2014shengbte}, TDEP\cite{tdep_iter}, D3Q-Thermal2,\cite{paulatto_anharmonic_2013, fugallo_ab_2013} almaBTE,\cite{carrete_almabte_2017} ALAMODE,\cite{tadano_anharmonic_2014} and $\kappa$ALDo.\cite{barbalinardo2020efficient}

As for the solution of the BTE, Phono3py, almaBTE, and $\kappa$ALDo implement the direct inversion or Eq.~\ref{mfp}, ShengBTE and TDEP use the iterative self-consistent approach, and Thermal2 the variational method.

All these codes have the option to read or calculate finite differences IFCs, except D3Q-Thermal2 which is fully integrated with the Quantum-Espresso suite,\cite{giannozzi_advanced_2017} and exploits density functional perturbation theory (DFPT) to compute both second and third-order IFCs. 
ShengBTE and AlmaBTE can read second-order IFCs computed by both finite differences and DFPT, while third-order IFCs are normally computed by finite differences.  
$\kappa$ALDo features full compatibility with Quantum-Espresso and D3Q force constants, finite-difference calculators, and interfaces with several force calculators, including LAMMPS and TDEP.

\section{Calculation Workflow}
\label{Sec:Workflow}

\begin{figure}[htb]
    \includegraphics[width=0.9\textwidth]{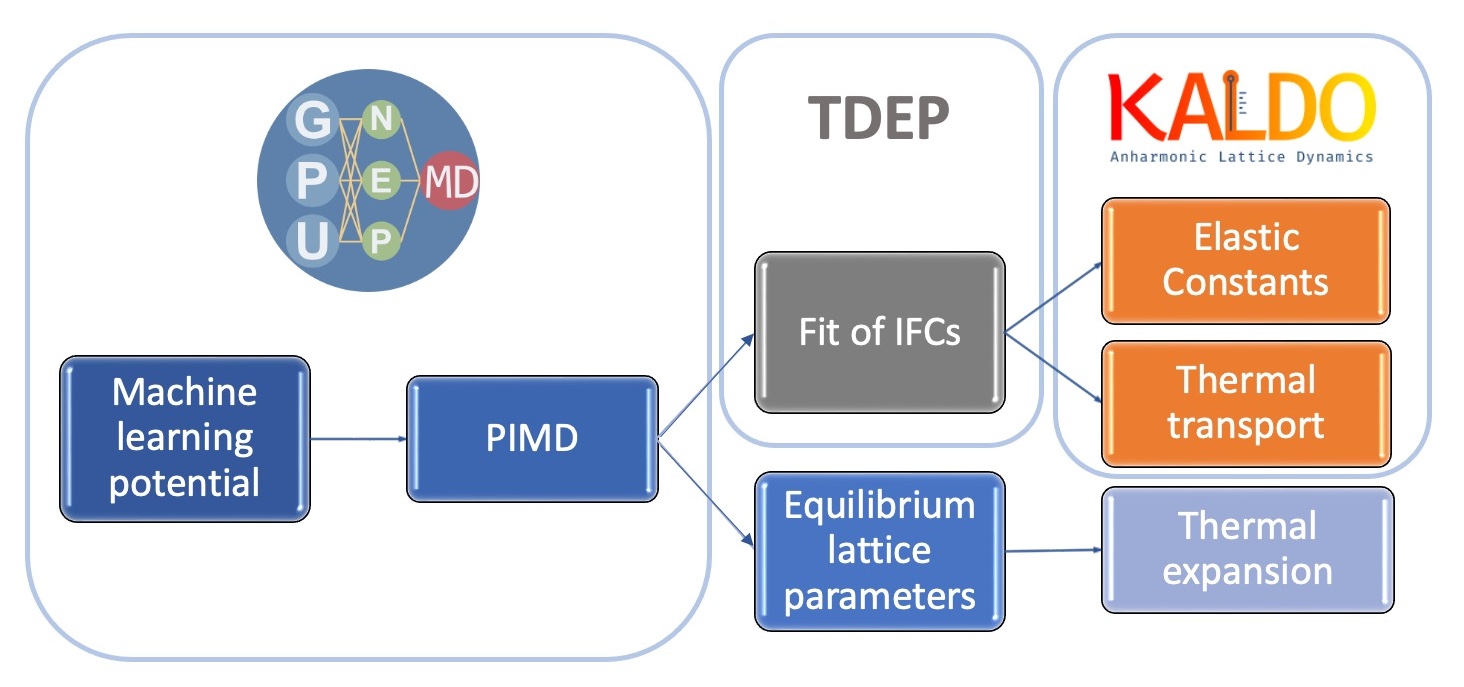}
\caption{Flowchart outlining the general process for calculating thermal and elastic properties by anharmonic lattice dynamics with temperature-renormalized force constants computed with TDEP from path integral molecular dynamics with machine learning potentials.}
    \label{fig:flowchart}
\end{figure}

The procedure proposed in this Article for calculating the elastic and thermal properties of materials at finite temperatures is outlined in the flowchart in Fig.~\ref{fig:flowchart}. 
The proposed procedure consists of the following steps: 
\begin{enumerate}
    \item Identify or fit from scratch a suitable MLP for the system of interest, in our case bulk silicon. Here we use a pre-trained NEP that was formerly shown to reproduce the vibrational and thermal properties of bulk crystalline silicon.\cite{fan2021neuroevolution} If necessary, NEP fitting can be performed with GPUMD.\cite{fan2022gpumd}
    \item Generate a sufficiently large supercell replicating the unit cell of the crystal of interest and run PIMD. We suggest running a shorter constant pressure simulation for equilibration and eventually a longer production run at constant volume for sampling displacements and forces in the canonical ensemble.
    \item Constant pressure PIMD simulations give the equilibrium lattice parameter at each temperature and allow one to infer directly the thermal expansion coefficients.
    \item Fit second- and third-order IFCs using TDEP.\cite{knoop2024tdep}
    \item Calculate the elastic constants from the second-order IFCs.
    \item Calculate the thermal conductivity in the ALD-BTE framework. For these last two steps, we decided to use $\kappa$ALDo, which is our in-house lattice dynamics code.
\end{enumerate}
We point out that some of these choices are arbitrary, for example, any combination of MLPs and codes can be used instead of NEP and GPUMD to perform finite temperature quantum sampling. Our choices are motivated by convenience and computational efficiency. 

\subsection{Convergence tests}

Each step of the workflow requires careful evaluations of the numerical results and convergence tests. The following parameters need to be considered: accuracy of the MLP, number of the PIMD beads, size of the supercell and spacial range of the TDEP fit, PIMD sampling time, Brillouin zone sampling (number of q-points) for the calculations of phonon properties, elastic constants, and thermal conductivity. 
Some of these checks may be performed for a single step, but for others, it is necessary to complete the full workflow to verify convergence.

\begin{figure}[t!]
    \centering
    \includegraphics[width=10cm]{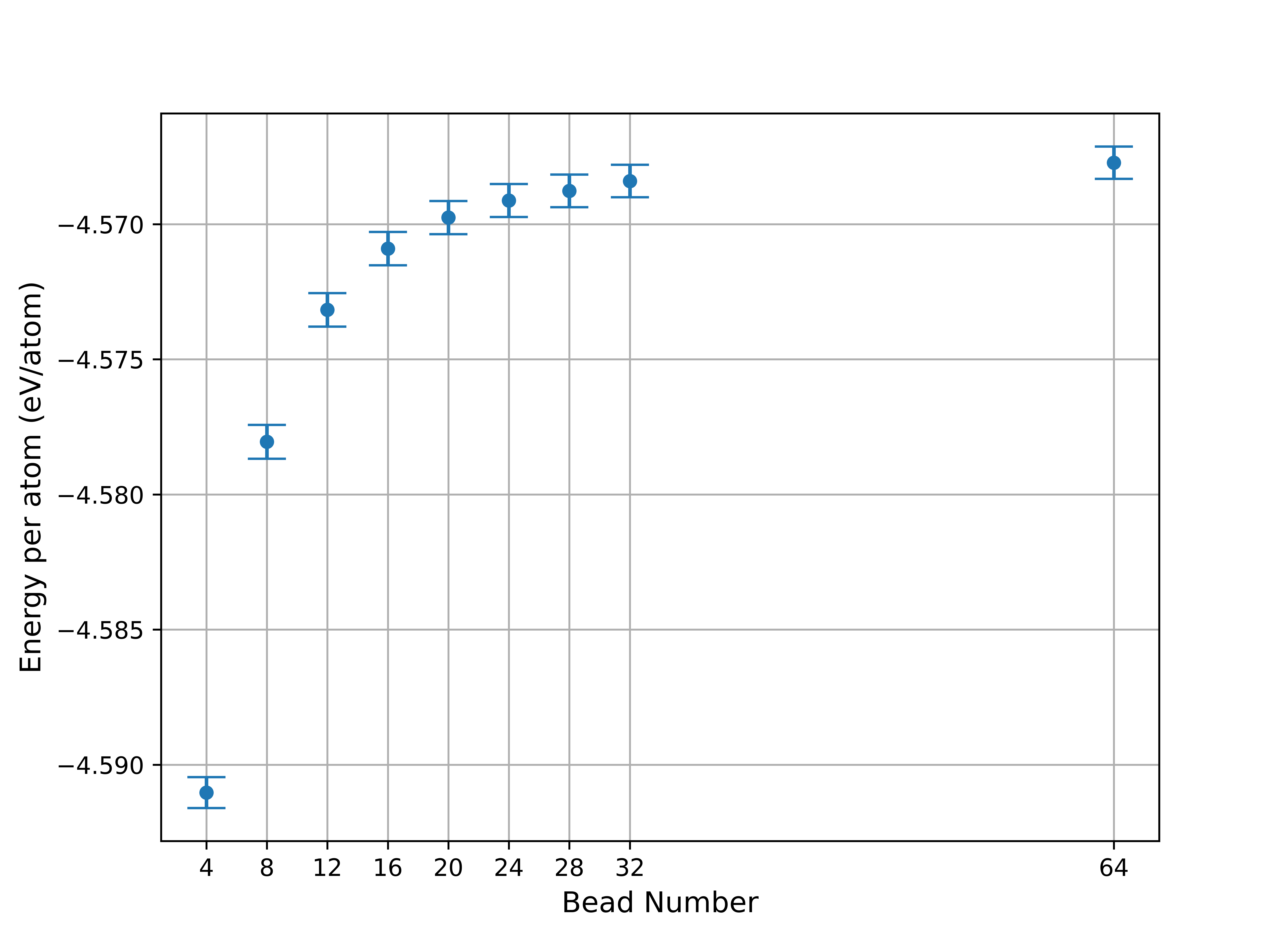}
    \caption{Plot of the total energy with respect to the number of beads at 50~K. Error bars indicate one standard deviation away from the average. $n=24$ can be considered sufficiently well-converged.}
    \label{fig:beadcon}
    \centering
\end{figure}
Here, we assume that the NEP model is sufficiently accurate as it was developed and tested in a previous work.\cite{fan2021neuroevolution} 
As the lowest temperature exhibits the largest nuclear quantum effects, we test the convergence PIMD total energy with the number of beads, $n$, at 50~K, and we can safely assume that $n=24$ is sufficiently well-converged at 50~K and higher temperatures (Fig.~\ref{fig:beadcon}). 

\begin{figure}[t!]
    \centering
    \includegraphics[width=10cm]{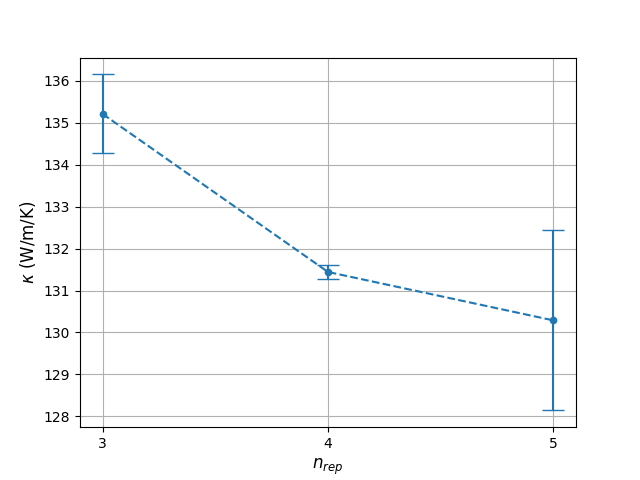}
    \caption{Plot of $\kappa$ vs $n_{rep}$ at 300~K where the conventional supercell is replicated $n_{rep}\times n_{rep} \times n_{rep}$ times.
    Error bars indicate one standard deviation away from the calculated value.}
    \label{fig:supcon}
    \centering
\end{figure}
To compute the IFCs by TDEP, we need to run PIMD on a supercell, i.e. a replica of the unit cell. The second-order IFCs are computed up to a cutoff radius which can be at most the largest sphere that can be included in the supercell. Third-order cutoffs were chosen such that the coefficient of determination is maximized, indicating the best fit. 
To obtain the largest cutoff radius with the smallest number of atoms it is convenient to use supercells as close as possible to cubic. In the case of bulk silicon, we can refer to the 8-atom conventional cubic cell as the unit cell and use $n_{rep}\times n_{rep}\times n_{rep}$ replicas of it. The convergence of $\kappa$ with $n_{rep}$ needs to be tested, which unfortunately implies running the full workflow in Figure~\ref{fig:flowchart}. IFCs are computed up to the maximum radius available within the supercell, corresponding to half the length of the supercell edge. 
$\kappa$ as a function of the size of the supercell at $T=300$~K is shown in Figure~\ref{fig:supcon}, with error bars computed by block averages over 5 blocks of the PIMD trajectories. 
$n_{rep}= 3$ already provides a reasonable estimate of $\kappa$, which can be well converged for $n_{rep}=4$ as the value is well within one standard deviation compared to $n_{rep}=5$. We then use $n_{rep}=4$ for the calculations at the other temperatures. 
The second-order IFC cutoff corresponds to twice the lattice parameter ($\sim 11$~\AA), while the optimal third-order IFC cutoff is approximately 6~\AA.
We note that $\kappa$ for silicon is not very sensitive to the range of the IFCs and the size of the supercell, as also observed in previous works.\cite{mcgaughey2019phonon} This would be different for polar materials where long-range interactions have a large impact on the thermal conductivity.\cite{ohtori_calculations_2009,ju_revisiting_2018,chen_strongly_2019}

\begin{figure}[bth]
    \centering
    \includegraphics[width=10cm]{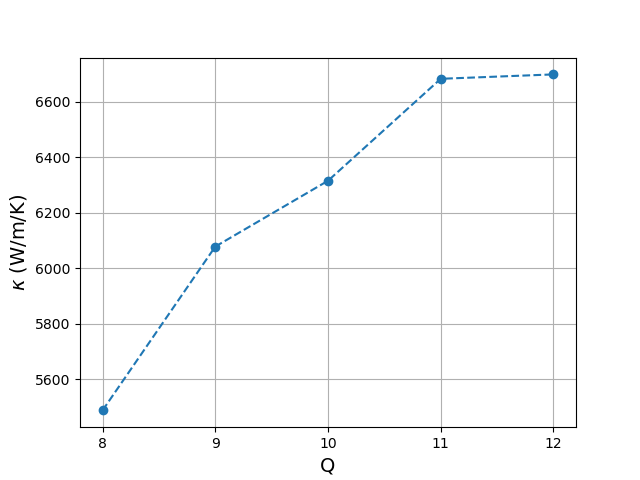}
    \caption{Plot of the thermal conductivity with respect to the number of q-points in each direction such that our q-grid is $Q\times Q\times Q$.}
    \label{fig:qcon}
    \centering
\end{figure}
Finally, we need to test the convergence of the thermal conductivity against the density of the q-point mesh used to sample the first Brillouin zone. We conducted this test at the lowest temperature considered, 50~K, as converging $\kappa$ at low temperature is usually more challenging. Figure~\ref{fig:qcon} shows that $\kappa$ plateaus for a  $11\times 11\times 11$ q-point grid. However, we choose to use a  $12\times 12\times 12$ grid since even grids are often computationally preferable for cubic systems. Note that this test was also conducted using the conventional 8-atom cell. The $12\times 12\times 12$ grid corresponds to a $19\times 19\times 19$ grid for the 2-atom primitive cell, which in our implementation of the BTE is well converged.\cite{barbalinardo2020efficient} 

\subsection{Thermal Expansion}

\begin{figure}[t!]
    \centering
    \includegraphics[width=10cm]{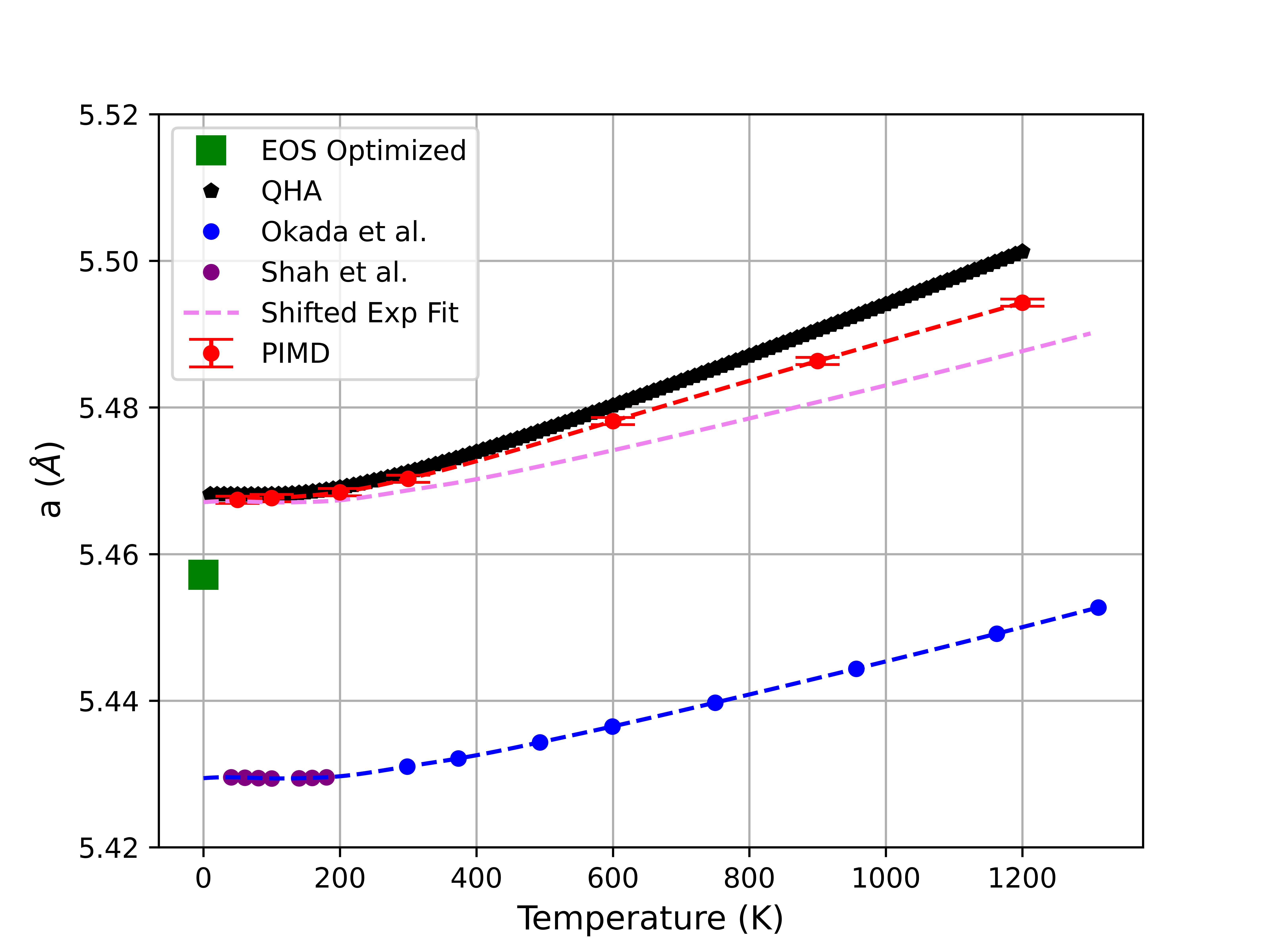}
    \caption{Lattice constant as a function of temperature, from constant pressure PIMD simulations (red dots), quasi-harmonic approximation (black dots), and experiments (blue and purple dots).\cite{okada1984precise,alpha_shah} The green square corresponds to the equilibrium lattice parameter from the equation of state at zero temperature without vibrational corrections. Calculated and experimental lattice parameters have been interpolated with cubic splines (red and blue dashed lines). The experimental curve has been shifted to align with the zero temperature limit of the calculations to facilitate comparison between theory and experimental trends. 
    }
    \label{fig:latticeconstant}
    \centering
\end{figure}

In the standard approach to computing $\kappa$ by first principles BTE, thermal expansion is not considered.\cite{broido_intrinsic_2007} 
In principle, one should calculate the lattice parameter and then the IFCs at each temperature to obtain a more accurate estimate of $\kappa$.
As described in section 2E, the QHA approach allows one to calculate the temperature-dependent lattice parameters at the cost of a few more phonon calculations. Here we compare the accuracy of QHA to the direct calculation of the lattice parameter of bulk silicon by constant-pressure PIMD. 
Figure~\ref{fig:latticeconstant} provides a comparison of these two approaches with the experimental measurements.\cite{okada1984precise, alpha_shah} 
The temperature dependence of the lattice parameter agrees with former quantum-mechanical calculations and experiments.\cite{therm_exp_kim_2017} 
It also reports the zero-temperature lattice constant computed from the equation of state. Quantum zero-point motion causes the shift from the zero-temperature lattice parameter and the limit of the QHA and PIMD simulations, which should coincide for $T\rightarrow 0$.  
This correction corresponds to 0.2$\%$ of the lattice parameter, in line with previous calculations.\cite{functionalzple}
For this system, QHA provides estimates of the lattice parameter consistent with PIMD up to room temperature. QHA and PIMD results start to diverge at temperatures larger than 600~K.

Theoretical results suffer from the systematic overestimate of the lattice parameter of the underlying DFT functional used to fit the NEP model, a well-known shortcoming of GGA.\cite{lattice_parameter_DFT} However, it is meaningful to compare the relative thermal expansion correcting for this systematic shift. 
Figure~\ref{fig:latticeconstant} shows that PIMD simulations provide a better agreement with experiments than QHA at room temperature and higher, although thermal expansion is still overestimated in our simulations. 
It is worth pointing out that above the Debye temperature (645~K for silicon) one can use classical MD, which is much more computationally efficient, instead of PIMD. 

To calculate the linear thermal expansion coefficient ($\alpha_{L}$), we have interpolated the calculated temperature-dependent lattice parameters with a cubic spline and computed the derivative at room temperature. Compared to the experimentally reported value of $ 2.72 \times 10^{-6}$ K$^{-1}$, we obtain a larger $\alpha_{L}$ by PIMD ($4.03 \times 10^{-6}$ K$^{-1}$) and even larger with QHA ($ 4.59 \times10^{-6}$ K$^{-1}$). 
This indicates that constant pressure PIMD (or MD) provides a better estimate of $\alpha_{L}$ than QHA, but the main source of discrepancies with experiments comes from the machine-learning potential or, most likely from the parent DFT functional used to construct the training set.

\subsection{Elastic Properties}

\begin{table}[ht!]
\caption{Elastic constants ($c_{ij}$), Bulk modulus ($B$), Young's moduli ($E$) and Shear modulus ($G$) of silicon in GPa computed from finite temperature second-order IFCs obtained by PIMD+TDEP.
Modeling results are compared to experimental, room temperature elastic constant, bulk modulus,\cite{elasticexp} and Young's moduli.\cite{youngmoduli} 
DFT data are from Bartok et al.\cite{bartok2018machine} and were obtained by computing the stress tensor at finite deformation of the cell parameters around equilibrium at 0~K (strain). Zero-temperature elastic constants computed with NEP with the stress/strain method and from the IFCs are also reported for comparison.
Finite temperature elastic constants are corrected with the difference between zero-temperature calculations with the two different approaches, thus fixing size convergence issues in the Born long-wave approach. Uncertainties calculated as the standard deviation found through block averaging of the trajectories at 50 K and determined to be: $\sigma_{11}\approx 0.2$ GPa, $\sigma_{12}\approx 0.3$ GPa, $\sigma_{44}\approx 0.1$ GPa.
}
\label{tab:elastics}
\centering
\begin{tabular}{l c c c c c c c c}
 \toprule
 Temperature (K) & $c_{11}$ & $c_{12}$ & $c_{44}$ & B & $E_{<111>}$ & $E_{<110>}$ & $E_{<100>}$ & $G$ \\ [0.5ex] 
  \hline
   Exp (300 K)\cite{elasticexp} 
              & 165.6 & 63.9 & 79.5 & 97.8 & 187.7 & 169.0 & 130.0 & 68.0 \\
 \hline
  DFT (0 K, Strain)\cite{bartok2018machine} 
              & 153.3 & 56.3 & 72.2 & 88.6 & 170.3 & 155.4 & 123.0 & 62.7 \\
 \hline
 0 (Strain)   & 135.1 & 60.3 & 67.2 & 85.2 & 159.6 & 137.9 & 85.8  & 55.6 \\
 0  (IFC)     & 137.2 & 61.9 & 67.6 & 87.0 & 161.0 & 139.0 & 98.7  & 55.6 \\
 \hline
 50           & 135.1 & 61.2 & 66.0 & 85.8 & 157.6 & 136.3 & 96.9  & 54.4 \\
 100          & 135.3 & 61.3 & 66.0 & 86.0 & 157.7 & 136.4 & 97.1  & 54.4 \\
 200          & 134.6 & 61.0 & 65.6 & 85.6 & 156.7 & 135.6 & 96.5  & 54.1 \\
 300          & 133.6 & 59.4 & 64.3 & 84.1 & 153.7 & 134.1 & 97.0  & 53.4 \\
 600          & 129.5 & 59.4 & 60.2 & 82.8 & 145.3 & 127.0 & 92.2  & 50.1 \\
 900          & 123.0 & 59.2 & 55.6 & 80.5 & 135.5 & 117.7 & 84.5  & 46.1 \\
 1200         & 116.8 & 57.7 & 50.5 & 77.4 & 124.5 & 108.7 & 78.7  & 42.1 \\
 \hline
\bottomrule  
\end{tabular}
\end{table}

As elastic constants can be directly inferred from second-order IFCs using the Born long-wave method, the proposed PIMD+TDEP scheme provides a viable tool to compute the elastic properties of solids at finite temperatures alternative to molecular dynamics or Monte Carlo\cite{ray_elastic_1988,sprik_second-order_1984, shinoda_rapid_2004} and potentially more efficient. 
The elastic constants reported in Table~\ref{tab:elastics} are in reasonable agreement with the experiments overall. The main reason for the observed discrepancies with experiments is the approximations in the DFT functional, but there are also differences between the DFT reference and the NEP model, size convergence issues inherent to the Born long-wave method, and statistical uncertainties in the TDEP fit.
We notice that the reference DFT functional\cite{bartok2018machine} systematically underestimates the elastic constants compared to experiments (see first two rows of Table~\ref{tab:elastics}), which is not uncommon for GGA functionals. 
To compare DFT and NEP, we computed the elastic constants at zero temperature by directly fitting the stress/strain relation through calculations of the stress tensor of strained supercells.\cite{fan2022gpumd} The NEP model further underestimates $c_{11}$ and $c_{44}$ but gives a slightly higher $c_{12}$. 
These calculations also serve as a reference to check the accuracy and the convergence of the Born long-wave method. Computing the second-order IFC matrix by finite differences at zero temperature, we find that elastic constants converge for a cutoff radius corresponding to half the edge of a $5\times 5\times 5$ supercell. Finite temperature calculations are performed with this supercell size. 

The finite temperature elastic constants exhibit the expected dependence on the temperature, as the material gets mechanically softer as the temperature increases. Softening is more pronounced for the diagonal components of the elastic tensor ($c_{11}$ and $c_{44}$), whereas the off-diagonal $c_{12}$ has a weaker temperature dependence. 
The decrease of the elastic constants becomes more prominent above room temperature and the trends are reflected in the Bulk, Young, and Shear moduli also reported in the table.
It is worth noting that there is a cross-over between $c_{12}$ and $c_{44}$ at about 600~K, as the latter becomes smaller than the former. 
This determines a change of sign of the Cauchy pressure, defined as $c_{12}-c_{44}$ which turns from negative to positive. A negative Cauchy pressure is characteristic of covalent directional bonding, whereas a positive value is typical of metallic bonding. 
This also marks a transition from brittle to ductile that takes place when the Pugh ratio $B/G$ exceeds 1.75.\cite{Pugh1954} In our calculations, this happens at 900~K. However, whereas this trend is physically meaningful, the NEP model's error on the Shear modulus is larger than that over the Bulk modulus, thus leading to an overestimate of the ductility of the material. More quantitative estimates of the elastic moduli would require better MLPs, possibly trained on datasets with a higher level of theory.

\subsection{Temperature Dependent Phonon Frequencies}
\begin{figure}[h]
    \centering
    \includegraphics[width=10cm]{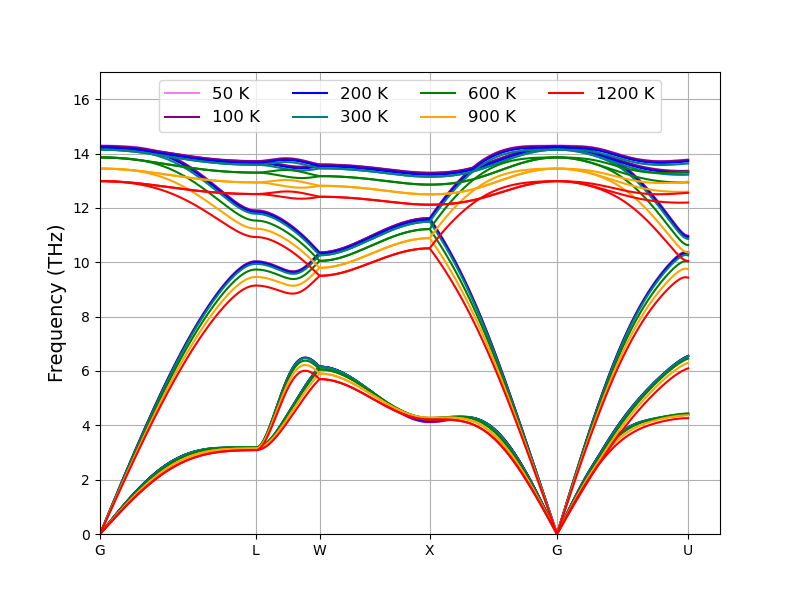}
    \caption{Dispersion relations of bulk silicon (primitive 2-atom cell) along high-symmetry directions in the Brillouin zone.
    } 
    \label{fig:dispersion}
\end{figure}

Anharmonicity in the potential causes a temperature renormalization of phonon frequency which affects the thermal properties of materials and stands at the base of experimental techniques such as Raman thermometry. 
Our workflow allows the calculation of phonon dispersions at finite temperatures (Figure~\ref{fig:dispersion}). Temperature renormalization leads to a monotonic shift toward lower frequencies and mostly affects optical modes. This shift can be substantial, exceeding 1~THz at the highest temperature (1200~K). 

\begin{figure}[t!]
    \centering
    \includegraphics[width=10cm]{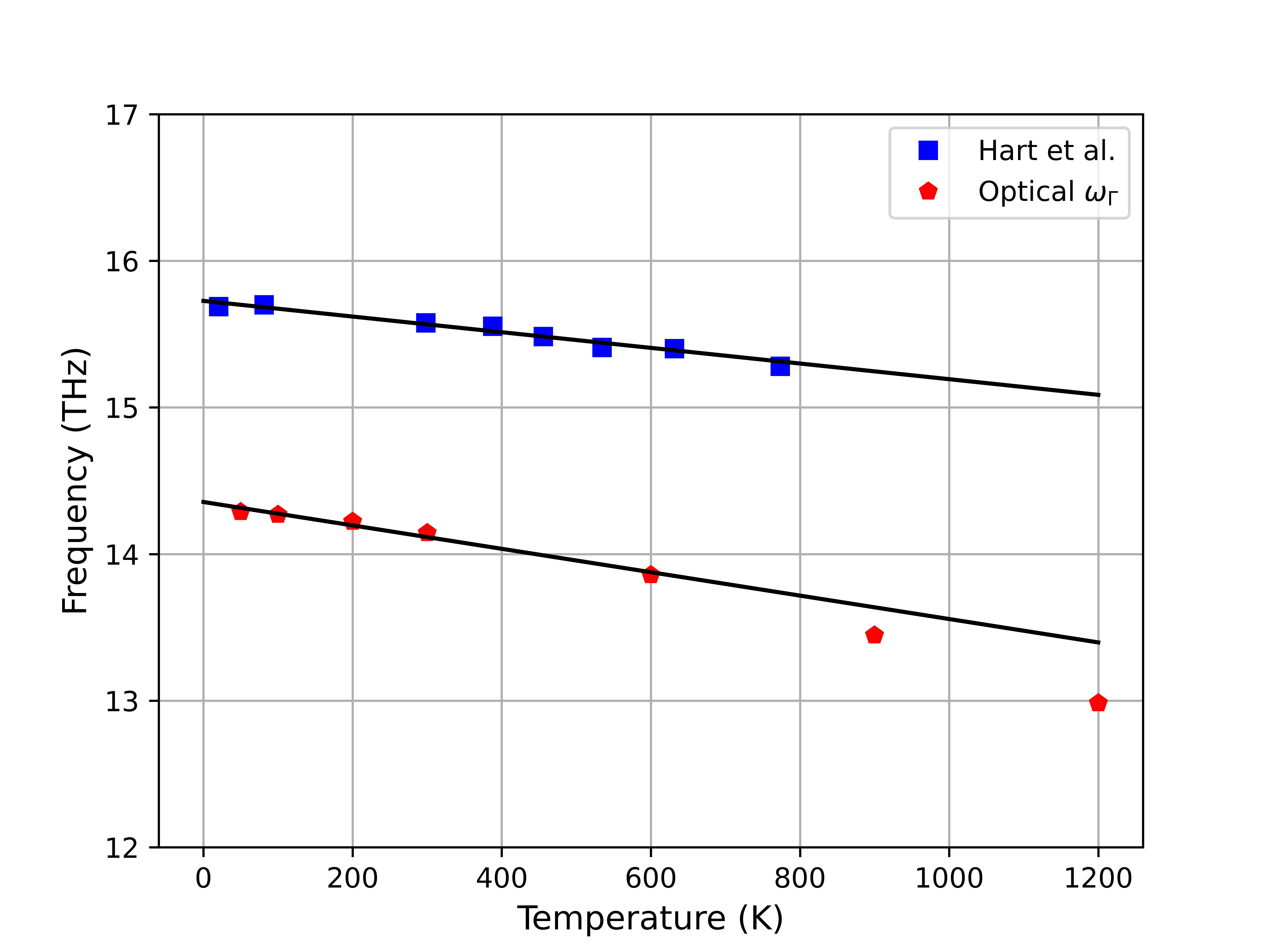}
    \caption{Finite temperature experimental Raman shift\cite{hart_raman_1970} and TDEP optical mode frequency at the \(\Gamma\)-point from $50-1200$ K. Line provided to guide readers visually and only fitted using data for \( T \leq 600 \; K\). Slope of fits are found to be $m_{TDEP} = -7.98\times10^{-4}\pm 4.58\times10^{-9} THz \  K^{-1}$ and $m_{EXP} = -5.34\times10^{-4}\pm 4.01\times10^{-9} THz \ K^{-1}$.}
    \label{fig:raman_shift}
\end{figure}
Temperature-dependent Raman measurements allow us to assess the accuracy of our calculations. In Figure~\ref{fig:raman_shift}, we compare the temperature dependence of the Raman peak to the frequency of the longitudinal optical mode at the center of the Brillouin zone. The absolute difference between the two data sets stems from approximations in the DFT functional and further inaccuracy in the NEP model. However, the temperature trends are in good agreement, with a slightly stronger temperature dependence in the calculations at temperatures higher than 300~K. {It is worth noting that, in the framework of TDEP with MD, the eigenvalues of the mass-normalized second-order IFC matrix correspond to the finite-temperature frequency spectrum.\cite{castellano2023mode} This is not exactly the case when sampling a Gaussian distribution as in the sTDEP or SCHA approaches.\cite{sscha_monacelli,castellano2023mode}}

\begin{table}[htb]
\caption{Speed of sound computed as the center-zone limit of the acoustic phonons group velocities $c\equiv\left|\lim_{\mathbf{q}\rightarrow 0}\nabla_{\mathbf{q}}\omega(\mathbf{q})\right|$ of the longitudinal (LA) and transverse acoustic (TA) modes along the $\Gamma-L$ and $\Gamma-X$ direction.
}
\label{tab:speedsound}
\centering
\begin{tabular}{c c c c c}
 \toprule
 Temperature (K) & \(c_{LA}^{\Gamma L}\) (\(\text{\AA}/ps\)) & \(c_{TA}^{\Gamma L}\) (\(\text{\AA}/ps\)) & \(c_{LA}^{\Gamma X}\) (\(\text{\AA}/ps\)) & \(c_{TA}^{\Gamma X}\) (\(\text{\AA}/ps\))\\ [0.5ex] 
  \hline
 50 & 87.25 & 45.29 & 82.78 & 53.15\\
 100 & 87.28 & 45.20 & 82.81 & 53.08\\
 200 & 86.98 & 45.16 & 82.54 & 52.90\\
 300 & 86.62 & 44.84 & 82.21 & 52.50\\
 600 & 84.72 & 43.72 & 80.46 & 50.97\\
 900 & 82.45 & 41.97 & 78.38 & 48.80\\
 1200 & 80.09 & 40.01 & 76.21 & 46.51\\
\bottomrule  
\end{tabular}
\end{table}

Temperature renormalization of acoustic modes is more subtle but may have a significant impact on the mechanical and thermal properties of materials. For example, the decrease of the elastic constants discussed in the previous section is strictly related to the softening of acoustic modes with temperature. This effect can be quantified by computing the group velocities of the three acoustic branches at the $\Gamma$-point corresponding to the speed of sound in the material. 
Table~\ref{tab:speedsound} shows that the group velocities of both longitudinal and transverse acoustic modes are considerably reduced at high temperatures (900, and 1200~K).


\subsection{Thermal Conductivity}


\begin{figure}[t!]
    \centering
    \includegraphics[width=10cm]{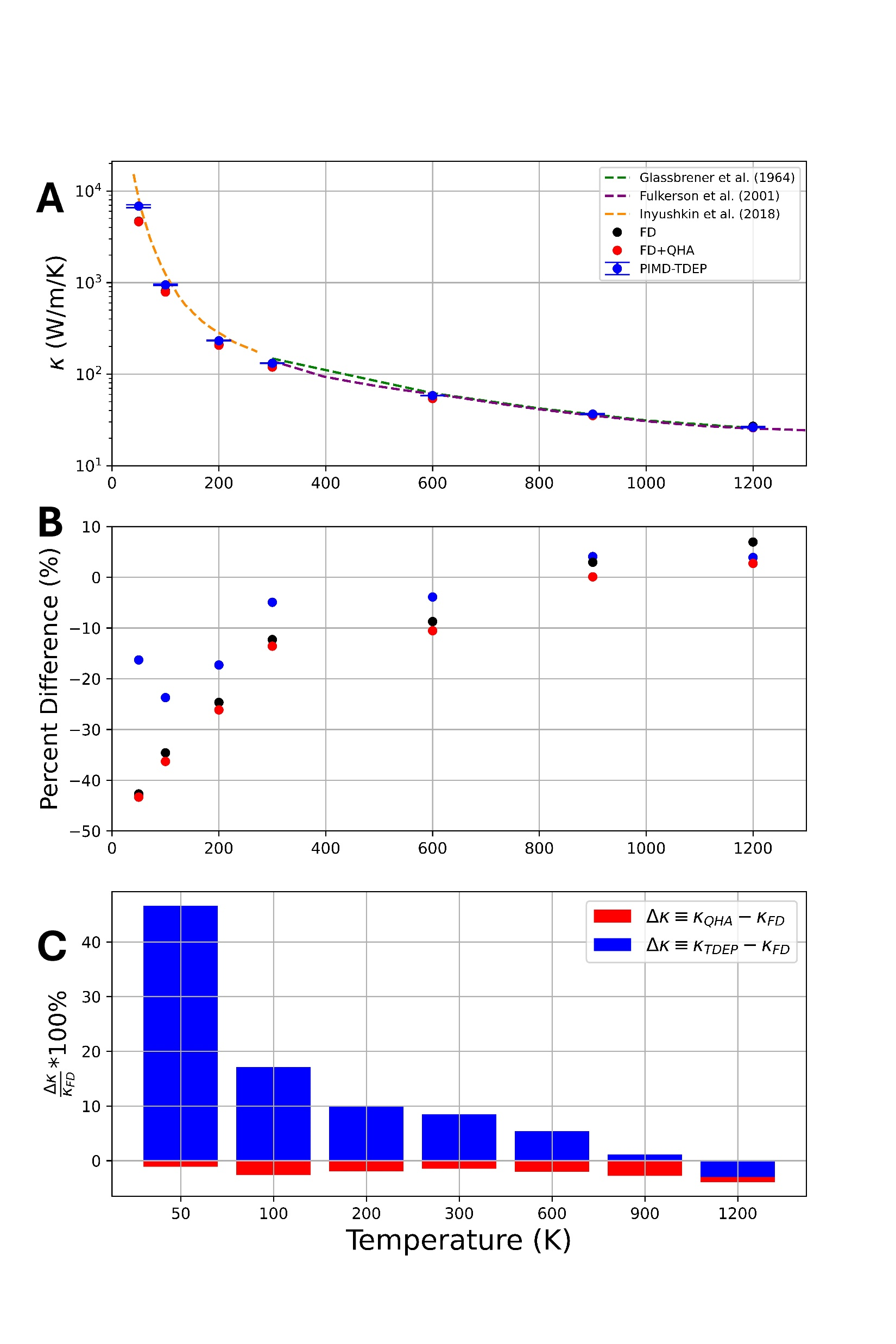}
    \caption{(a) Experimental and calculated $\kappa$ as a function of temperature. (b) Percent difference of $\kappa$ with respect to experimental\cite{kappa_glass, kappa_fulk, kappa_inyush} measurements from MD-TDEP, finite-different and finite-difference with QHA methods. (c) Percent correction on $\kappa$ as a function of temperature from finite-difference compared to finite-difference with QHA ($\Delta\kappa = \kappa_{QHA} - \kappa_{FD}$) and finite-difference compared to TDEP methods ($\Delta\kappa = \kappa_{TDEP} - \kappa_{FD}$).}
    \label{fig:kappa}
\end{figure}



\begin{figure*}[t!]
\centering 
\includegraphics[width=.49\linewidth]{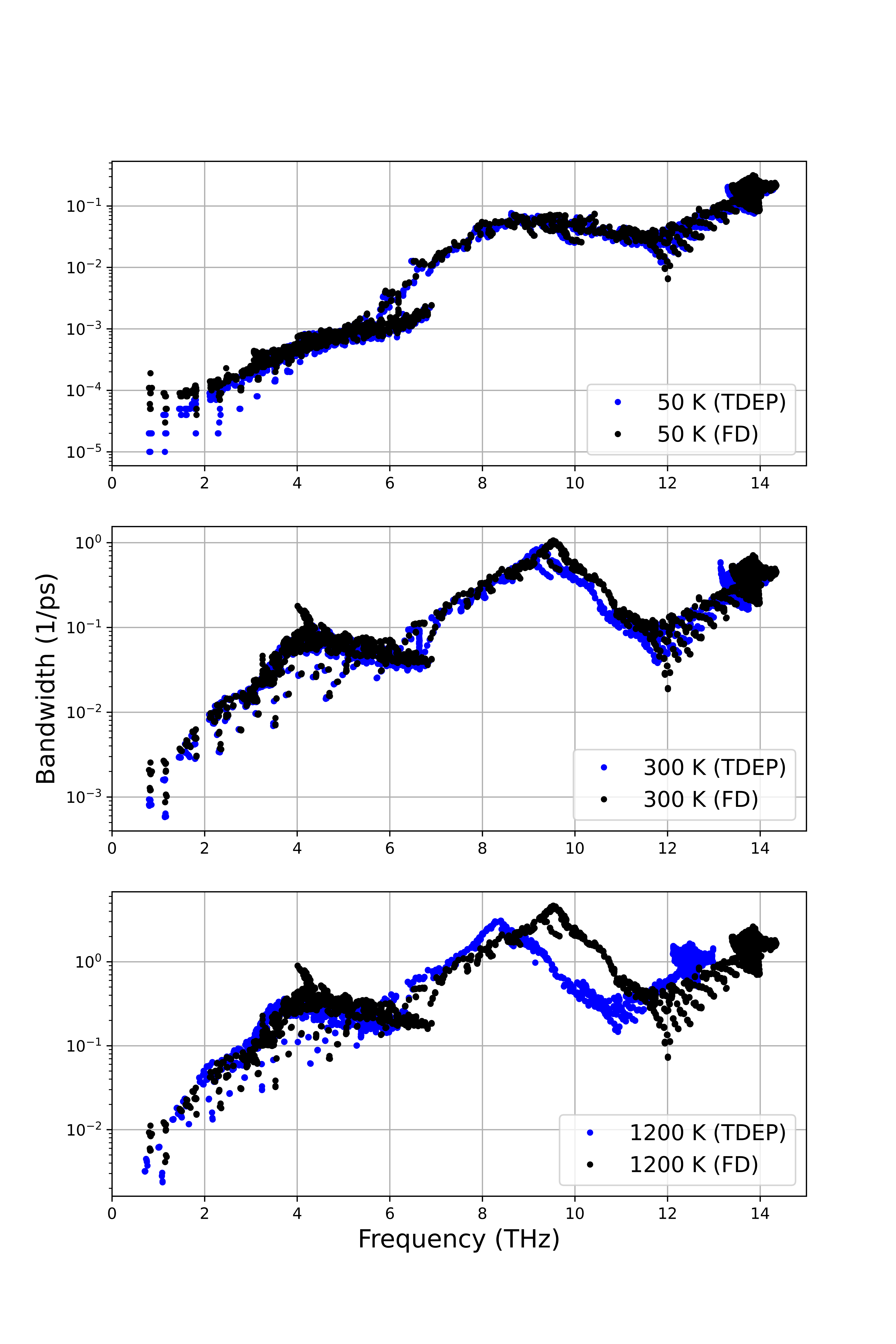}
\includegraphics[width=.49\linewidth]{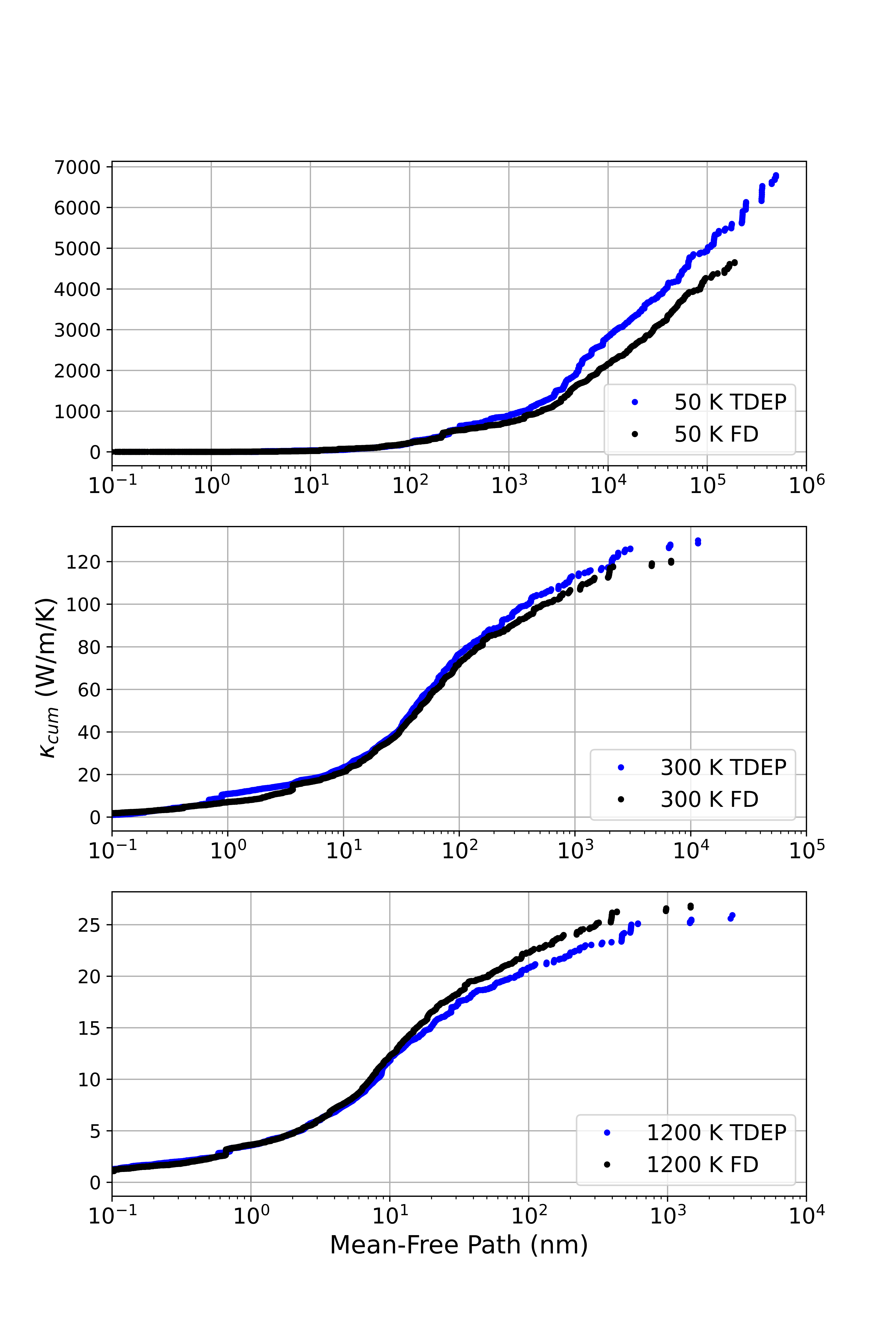}

\caption{Left panels: Phonon bandwidth computed using finite-temperature force constants (blue) compared to finite differences bandwidth (black) at 50 K, 300 K, and 1200 K. 
 Right panels: Cumulative thermal conductivity vs. mean free path  $\kappa_{cum}\left(\lambda\right)\equiv\int_{0}^{\lambda}\frac{1}{N}\sum_{\mu}\kappa_{\mu}\delta(\lambda_{\mu}-\lambda')d\lambda'$ 
 from the Boltzmann transport equation with IFCs computed by finite differences (FD) and TDEP.}
\label{fig:gamma_vs_freq}
\end{figure*}


\mycomment{
\begin{figure*}[t!]
\centering 
\includegraphics[width=.49\linewidth]{k_cum.png}
\caption{
 Cumulative thermal conductivity vs. mean free path  $\kappa_{cum}\equiv\int_{0}^{\lambda}\frac{1}{N}\sum_{\mu}\kappa_{\mu}\delta(\lambda_{\mu}-\lambda')d\lambda'$ 
 from the Boltzmann transport equation with IFCs computed by finite differences (FD) and TDEP. }
\label{fig:mfp_vs_freq}
\end{figure*}
}

The main goal of this study is to develop and assess a workflow to calculate thermal conductivity including temperature effects more accurately than in standard anharmonic lattice dynamics approaches.   
$\kappa(T)$ computed by inverting the BTE with second and third-order IFCs obtained from the PIMD+TDEP workflow is plotted in Figure~\ref{fig:kappa} alongside experimental measurements.\cite{kappa_fulk,kappa_glass,kappa_inyush}  
Our results are compared to calculations performed with IFCs obtained by finite differences (FD) at zero temperature either with the lattice parameter corresponding to the minimum of the equation of state or considering thermal expansion in the QHA. 
Figure~\ref{fig:kappa}A shows that PIMD+TDEP results are closer to the experimental reference, especially at low temperatures. The percent difference between calculations and experiments is reported in Figure~\ref{fig:kappa}B. Including nuclear quantum effects in the calculation of both second and third-order IFCs largely improves the accuracy of low-temperature calculations. Conversely, for silicon, the renormalization of the IFCs at temperatures above 600~K has a relatively smaller effect. 
The correction produced by computing IFCs with PIMD+TDEP compared to FD (Figure~\ref{fig:kappa}C) is positive, except for $T=1200$~K and the largest at the lowest temperatures. The QHA correction, in turn, is always negative and remains similar at all temperatures. 

To analyze the source of the differences between PIMD+TDEP and FD results across the broad temperature range considered, we computed anharmonic phonon properties, including phonon linewidths (which are the inverse of phonon lifetimes), and the cumulative thermal conductivity as a function of phonon mean free paths 
 at 50~K, 300~K, and 1200~K (Figure~\ref{fig:gamma_vs_freq}). 
Phonon linewidths provide a measure of the anharmonicity of vibrational states, and mean free paths, together with phonon group velocities, provide phonon-resolved contributions to the total lattice thermal conductivity, according to Equation~\ref{kappa}.  
In general, we observe that the linewidths of low-frequency acoustic modes computed from PIMD+TDEP are significantly smaller than those obtained from FD. 
Up to 300~K, we do not see significant temperature effects on phonon group velocities, hence we can conclude that the large thermal conductivity corrections at low temperatures stem from nuclear quantum effects on the anharmonic phonon properties, i.e. third-order IFCs. Indeed, we observe substantially longer phonon mean free paths providing a large contribution to $\kappa$ at 50~K. 
At room temperature, PIMD+TDEP and FD give similar $\kappa$. The slightly larger $\kappa$ obtained by PIMD+TDEP is due to a slightly lower anharmonicity of the low-frequency acoustic modes that exhibit longer lifetimes leading to longer mean free paths than those predicted by FD. 
At 1200~K, $\kappa$ from TDEP+PIMD is lower than from FD. This is mostly due to the renormalization of phonon group velocities as anharmonic effects make acoustic modes softer, whereas lifetimes computed from temperature-renormalized IFCs still result slightly longer than from FD. It should be stressed that four-phonon and higher-order scattering processes become more impactful at high temperatures and would further reduce the thermal conductivity, potentially bringing it more in line with experiment\cite{4ph_si_2017}.

Silicon is usually considered a simple benchmark system for thermal conductivity calculations, as it is made of relatively heavy atoms and does not exhibit temperature-stabilized phases. In fact, the standard FD approach usually provides reasonable results, especially around room temperature.\cite{broido_intrinsic_2007,mcgaughey2019phonon}
Instead, we found that accounting for nuclear quantum effects on third-order IFCs is essential to compute $\kappa$ at low temperatures and provides a substantial correction to $\kappa$ even at room temperature. This is probably the case for several other crystalline solids with high Debye temperatures that are weakly anharmonic and commonly considered suitable for treating with a perturbative FD approach. Generally, we see that the temperature renormalization of third-order IFCs produces a positive correction of $\kappa$. 
As expected, at high temperatures, above 600 K, frequency renormalization plays a role and results in a decrease of phonon group velocities, thus providing a negative correction to $\kappa$. In our silicon model, the former effect is predominant up to 600~K, whereas the latter becomes predominant at 1200~K.

\section{Conclusions}
In summary, we introduced a workflow that uses PIMD and TDEP to account for nuclear quantum effects and temperature renormalization of the interatomic force constants, providing a more accurate way to compute the elastic and thermal properties of materials at finite temperatures. 
We have illustrated the theory and the step-by-step procedure, including convergence tests and finite-size scaling, to apply this approach to a benchmark crystalline system, bulk silicon. 
The results highlight the importance of considering nuclear quantum effects in the calculation of anharmonic force constants to obtain an accurate estimate of the thermal conductivity at low temperatures. 
These modern numerical methods, when combined, achieve closer alignment with experimental results providing a basis for studying more novel or poorly understood materials that exhibit temperature-induced phase transitions, strong anharmonicity, or substantial nuclear quantum effects.


\section*{Data Availability}
The complete input files of PIMD simulations, TDEP force constant fitting, BTE calculations with $\kappa$ALDo, alongside the supplement data are publicly available at \href{https://github.com/dafolkner/Silicon_project}{\textcolor{red}{\small https://github.com/dafolkner//Silicon\textunderscore project}}.

\begin{acknowledgments}
We are grateful to Olle Hellman and Matthieu J. Verstraete for fruitful discussions on theories and implementations of temperature-dependent effective potentials. 
This work was partially supported by the U.S. Department of Energy, Office of Basic Energy Sciences, Division of Materials Science and Engineering (Grant No. DE-SC0022288).
Florian Knoop acknowledges support from the Swedish Research Council (VR) program 2020-04630, and the Swedish e-Science Research Centre (SeRC).
\end{acknowledgments}

\bibliography{bibliography.bib}

\end{document}